\documentclass[3p,times]{elsarticle}

\usepackage{ecrc}

\volume{00}

\firstpage{1}

\journalname{Physics Procedia}

\runauth{Z. Salman et al.}

\jnltitlelogo{Physics Procedia}

\usepackage{amssymb}

\usepackage[figuresright]{rotating}

\newcommand{\bnmr}{$\beta$-NMR}
\newcommand{\Li}{$^8$Li}

\newcommand{\lemusr}{LE-$\mu$SR}

\begin{document}

\begin{frontmatter}
\dochead{}

\title{Proximal magnetometry of monolayers of magnetic moments}

\author[PSI]{Z. Salman\corref{cor}}
\ead{zaher.salman@psi.ch}
\author[OX]{S. J. Blundell}

\cortext[cor]{Tel. +41-56-310-5457}

\address[PSI]{Laboratory for Muon Spin Spectroscopy, Paul Scherrer
  Institut, CH-5 232 Villigen PSI, Switzerland}
\address[OX]{Clarendon Laboratory, Department of Physics, Oxford
  University, Parks Road, Oxford OX1 3PU, UK}

\begin{abstract}
  We present a method to measure the magnetic properties of monolayers
  and ultra-thin films of magnetic material. The method is based on
  low energy muon spin rotation and $\beta$-detected nuclear magnetic
  resonance measurements. A spin probe is used as a ``proximal''
  magnetometer by implanting it in the substrate, just below the
  magnetic material. We calculate the expected magnetic field
  distribution sensed by the probe and discuss its temperature and
  implantation depth dependencies. This method is highly suitable for
  measuring the magnetic properties of monolayers of single molecule
  magnets, but can also be extended to ultra-thin magnetic films.
\end{abstract}

\end{frontmatter}

\section{Introduction}
Recent developments of low energy muon spin rotation (LE-$\mu$SR)
\cite{Morenzoni94PRL,Prokscha08NIMA} and $\beta$-detected nuclear
magnetic resonance ($\beta$-NMR) \cite{Salman04PRB,Morris04PRL},
provide unique local spin probe tools for measurements in thin films
and multilayers. However, the application of these methods is limited
by a minimal thickness of films, which provides sufficient stopping
power for the implanted probes. For typical density materials a
minimal 1-2 nm thickness is required at the lowest available
implantation energy (1-2 keV). Therefore, these methods are generally
not suitable for studies of monolayers and ultra-thin films.
Nevertheless, for magnetic materials it is possible to perform a
``proximal'' measurement by implanting the probe in the substrate (or
an under-layer) of the material, and sensing the dipolar magnetic
fields from the layer of interest \cite{Salman07NL,Salman09ML}.

To date, measurements of the magnetic properties of monolayers of
single molecule magnets \cite{GatteschiMNN} (SMMs) have been mostly
performed using X-ray absorption spectroscopy (XAS) and X-ray magnetic
circular dichorism (XMCD) \cite{Mannini09NM,Mannini10N}.  These are
typically limited to high magnetic fields and provide information
regarding the average static magnetic properties of the irradiated
portion of the monolayer. In contrast, LE-$\mu$SR and $\beta$-NMR
provide local probe measurements with important advantages;
sensitivity to a large range of spin fluctuations and dynamics and
applicability in zero or any applied magnetic field. However, a
detailed interpretation of the measured spectra using LE-$\mu$SR
\cite{Salman09ML} or $\beta$-NMR \cite{Salman07NL} when implanting the
probe just below the monolayer is still absent. In this paper we
calculate the magnetic field distribution in a non-magnetic substrate
due to a monolayer of SMMs. This distribution can be used to fit
LE-$\mu$SR and \bnmr\ measurements, providing information regarding
the magnetic and geometric properties of the monolayer. The
calculations can be easily generalized for the case of ultra-thin
magnetic films by considering the domains as individual magnetic
moments with a given average size.

\section{Uniformly Magnetized Sheet Approximation}
\begin{figure}[h]
\begin{minipage}[t]{0.45\linewidth}
\includegraphics[width=0.9\columnwidth]{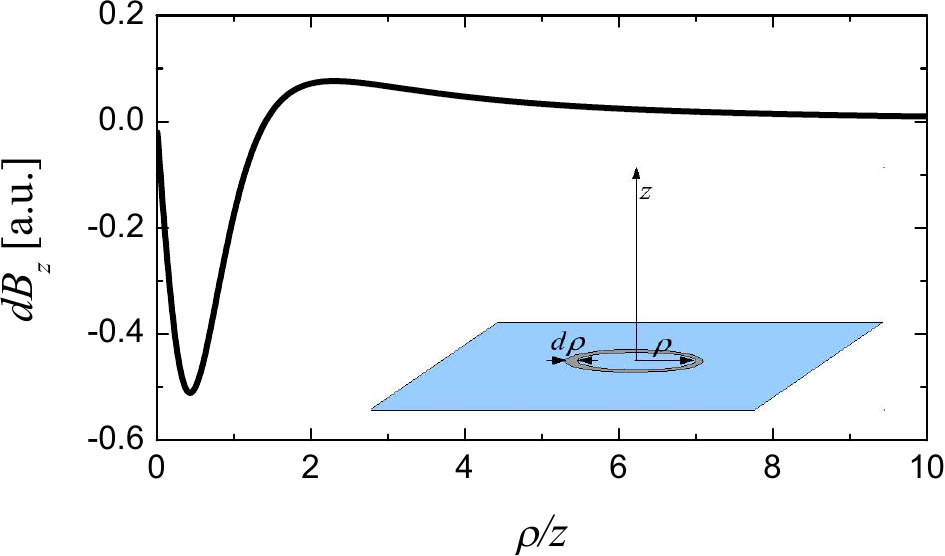}
\caption{The contribution $dB_{z}$ of one annular region as a function
  of $\rho/z$. The inset shows a uniformly magnetized sheet divided
  into annular regions.} \label{dBz}
\end{minipage}
\hspace{0.05 \linewidth}
\begin{minipage}[t]{0.45\linewidth}
  \centerline{\includegraphics[width=0.9\columnwidth]{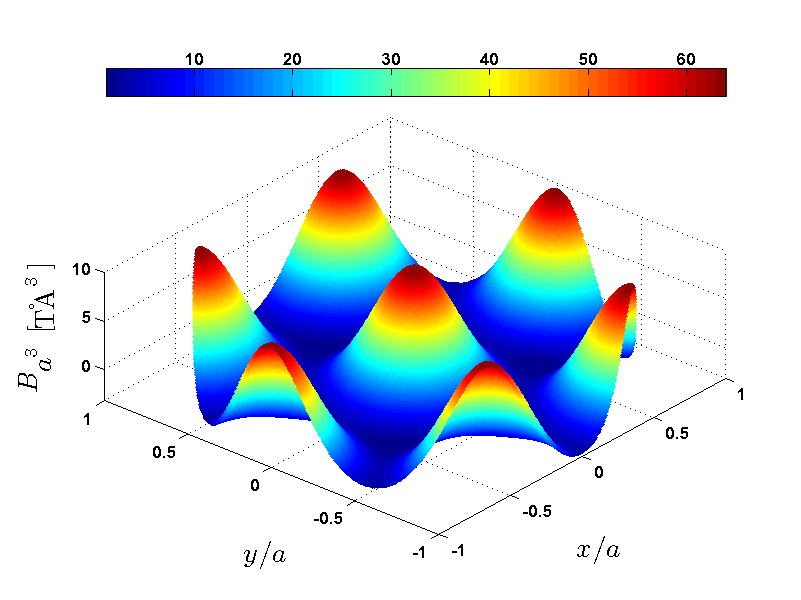}}
  \caption{The $z$ component of the field, sensed by a spin probe at
    $z=a/2$, as a function of position. The color scale represents the
    strength of the field.}
  \label{Layer}
\end{minipage}
\end{figure}

As a simple approximation, a monolayer of magnetic moments (or a thin
film) can be viewed as a uniformly magnetized sheet. Assuming magnetic
moment $M$ per unit area, located in the plane $z=0$ (see inset of
Fig.~\ref{dBz}) with the magnetization aligned along $-\hat{z}$. The
magnetic scalar potential from an annular region of radius $\rho$ and
width $d\rho$ is
\begin{equation}
  d\phi_{M}(\rho,z)=-\frac{Mz\rho d\rho}{2(z^{2}+\rho^{2})^{3/2}}.
\end{equation}
The contribution of such a region to the magnetic field at a distance
$z$ from its center is
\begin{equation} \label{dBzeq}
  dB_{z}(\rho,z)=-\mu_{0}\frac{\partial (d\phi_{M})}{\partial z}=\frac{\mu_{0}M}{2}\frac{\rho(\rho^{2}-2z^{2})}{(\rho^{2}+z^{2})^{5/2}}d\rho,
\end{equation}
where $\mu_0$ is the permeability of the vacuum. We plot this
contribution as a function of $\rho/z$ in Fig. \ref{dBz}.
Integrating the contributions from all annuli gives a magnetic field
zero as expected.

However, this approximation fails when looking at the microscopic
structure of the sheet, i.e., as a collection of individual magnetic
moments or even randomly oriented magnetic domains. This is the case
when measuring the field due to such sheet using a local spin probe,
which is positioned at a distance smaller than the microscopic details
of the sheet (distance between moments or size of domains). For
example, if we have a monolayer of moments with average spacing $a$
between them, then the magnetic field sensed by a spin probe at a
distance $z<a$ from the monolayer is non-zero.

\section{Monolayer of Magnetic Moments}
\begin{figure}[h]
  \centerline{\includegraphics[width=0.40\columnwidth]{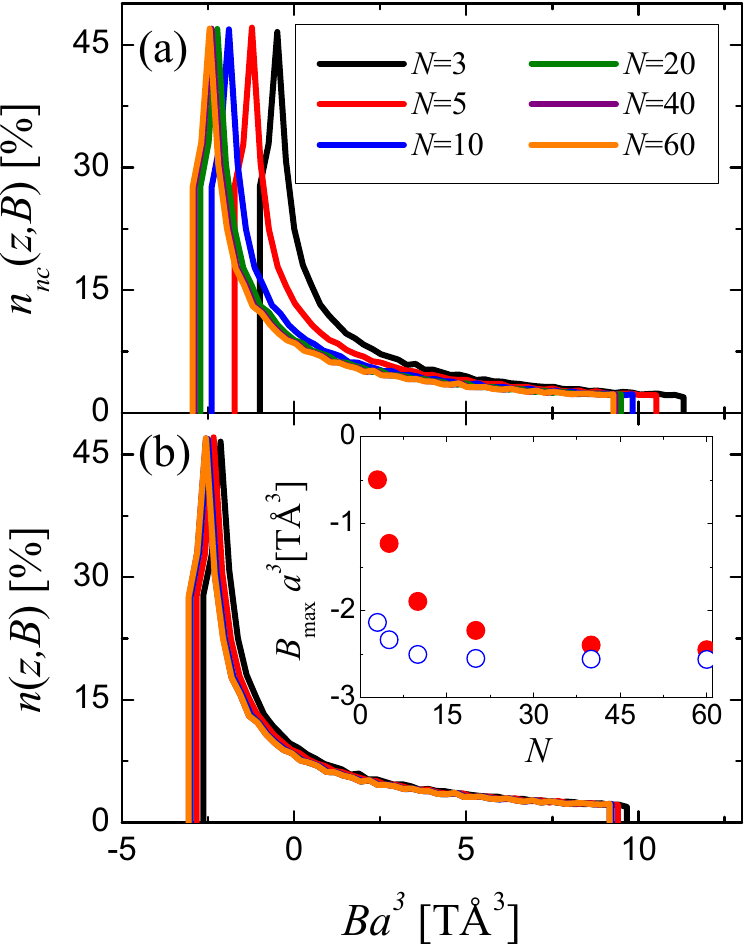}}
  \caption{The distribution of $B_z$ sensed by a spin probe due to a
    patch of moments with radius $N$ (a) before and (b) after the
    correction in Eq. \ref{correction}. The inset shows the most
    probable field as a function of $N$ before (full symbols) and
    after (empty symbols) correction.}
  \label{SumDip}
\end{figure}

In order to adequately account for the microscopic structure of a
monolayer of magnetic moments, $\vec{\mu}_i$, one needs to sum the
contribution of all moments in the monolayer. The dipolar field from
$\vec{\mu_i}$ at $\vec{r}_i=x_i \hat{x}+y_i \hat{y} + z_i \hat{z}$ is
\begin{equation}
  \vec{B}_i(\vec{r}_i)=\frac{\mu_0}{4\pi}\frac{3\vec{r}_i(\vec{\mu}_i\cdot\vec{r}_i)-\vec{\mu}_i r_i^{2}}{r_i^{5}}.
\end{equation}
A typical NMR or transverse field $\mu$SR measurement is sensitive to
the field distribution of one component of the field. In what follows
we will restrict ourselves to the $B_{z}$ component, where $z$ is
normal to the monolayer surface. Working in polar coordinates with
$\vec{\mu}_i=\mu_\rho^i \hat{\rho}+\mu_z^i \hat{z}$ and $\vec{r}_i=
\rho_i \hat{\rho}+z_i \hat{z}$ we can write
\begin{equation} \label{SingleMom}
  B_{z}^i(\rho_i,z_i)=\frac{\mu_{0}}{4\pi}\frac{3z_i\rho_i\mu_{\rho}^i+2z_i^{2}\mu_{z}^i-\rho_i^{2}\mu_{z}^i}{(\rho_i^{2}+z_i^{2})^{5/2}}.
\end{equation}
Therefore the field distribution, $n(z,B)$, can be obtained by summing
the contribution from all magnetic moments. For simplicity, we start
by considering a monolayer of moments $\vec{\mu}_i=\mu \hat{z}$ in the
plane $z=0$, arranged on a triangular lattice with lattice constant
$a$. For example, taking $\mu=\mu_B$, a spin probe at $z=a/2$ will
sense dipolar fields as shown in Fig.~\ref{Layer}. The resulting field
distribution is shown in Fig.~\ref{SumDip}(a), where different curves
are results of summation of a different number of moments, such that
$N$ is the radius (in units of $a$) of the considered ``patch''. In
order to sample a field distribution as in Fig.~\ref{Layer}, it is
sufficient to sample only a unit cell of an equilateral ($a$)
triangular area between 3 moments\footnote{Due to the symmetry of the
  lattice the sampling can also be done on 1/6 of this area}. Note
that the qualitative shape of the distribution is almost independent
of $N$. However, due to the finite number of considered moments, we
find a strong shift which decreases as a function of $N$.

In order to correct for this ``artifact'' we need to account for all
the moments in the infinite monolayer. However, since the moments
outside the patch satisfy the criterion of a uniformly magnetized
sheet, i.e. $r_i \gg a$, their contribution can be calculated as an
infinite sheet of uniform magnetization with a hole of radius
$\rho_0=aN$ around the origin. This can be calculated by integrating
$dB_z$ (Eq.~\ref{dBzeq}) between $\rho=\rho_0$ and infinity,
\begin{equation} \label{DeltaBz}
  \Delta B_{z}=\int_{\rho_{0}}^{\infty}dB_{z}(\rho,z) = \frac{\mu_0 M}{2 \rho_0 (1+\zeta^2)^{3/2}},
\end{equation}
with $\zeta=z/\rho_0$. In contrast to the case of a uniformly magnetized
sheet, $\Delta B_z$ here does not vanish for finite $\rho_0$. Note
that when the spin probe is far from the monolayer, such that
$\rho_{0}\gg z$, then $\Delta B_{z}\approx\frac{\mu_{0}M}{2\rho_{0}}$.
This is quite different from the three dimensional case, where the
remaining contribution is $\frac{\mu_{0}M}{3}$ (where in this case $M$
is the three-dimensional magnetization, i.e. the magnetic moment per
unit volume), independent of how far we integrate out to.

Assuming uniform magnetization due to moments $\mu=\mu_B$, ordered on
a triangular lattice with lattice constant $a$, the magnetic moment
per unit area is $M=\mu_{B}/(\sqrt{3}a^{2}/4)$. Using $N=\rho_0/a$ we
arrive at,
\begin{equation} \label{correction}
  \Delta B_{z}=\frac{2\mu_{0}\mu_{B}}{\sqrt{3}a^{3}N (1+\zeta^2)^{3/2}}.
\end{equation} 
This is the cause of shift seen in Fig.~\ref{SumDip}(a), and should be
subtracted from the dipolar field of the patch of moments to correct
for its finite size effect. Comparison between the field distribution
before (Fig.~\ref{SumDip}(a)) and after (Fig.~\ref{SumDip}(b)) the
correction shows clearly how well it works. Taking into consideration
a patch with $N=10$ is already sufficient to accurately calculate the
field distribution. Note that without the correction, even a patch
with $N=60$ exhibits a small artificial shift (inset of
Fig.~\ref{SumDip}).

The field distribution has a similar shape to that in the vortex state
of a superconductor \cite{Brandt88JLTP}. The high field cut-off is due
to probes stopping exactly below a magnetic moment, while the low
field cusp is due to probes stopping in middle between three
neighbouring moments. However, one important difference between this
case and a vortex lattice is that in the former the field distribution
depends strongly on the stopping depth of the probe while in a
superconductor the depth dependence is much weaker
\cite{Niedermayer99PRL,Salman07PRL}. As we show below, this strong
depth dependence will alter the shape of the field distribution
entirely in a real measurement. Nevertheless, it is important to point
out here that the shape of $n(z,B)$ is determined by the lattice
constant $a$ and and the size of magnetic moment $\mu$. Hence, these
parameters can, in principle, be extracted directly from a measurement
of the field distribution. Also, the details of the lattice, e.g.
triangular or other, do not alter the qualitative shape of $n(z,B)$
dramatically.

\section{Measured Field Distribution}
\begin{figure}[h]
\begin{minipage}[t]{0.45\linewidth}
  \centerline{\includegraphics[width=0.9\columnwidth]{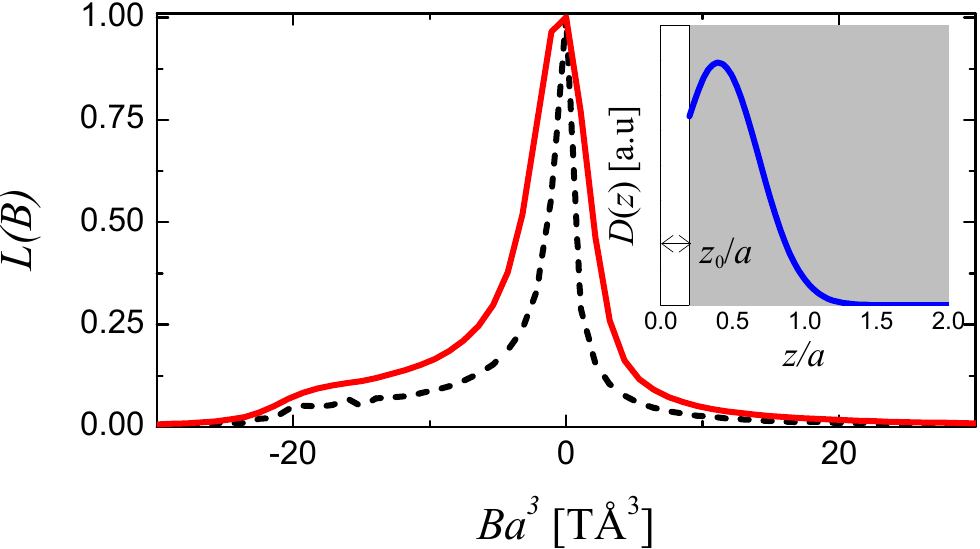}}
  \caption{The field distribution $L(B)$ sensed by a spin probe due to
    a monolayer of moments at $z=0$. The dashed and solid lines are
    calculated with and without broadening, respectively. The inset
    shows a typical stopping profile of implanted probes in the
    substrate (shaded area).}
  \label{Measured} 
\end{minipage}
\hspace{0.05 \linewidth}
\begin{minipage}[t]{0.45\linewidth}
  \centerline{\includegraphics[width=0.9\columnwidth]{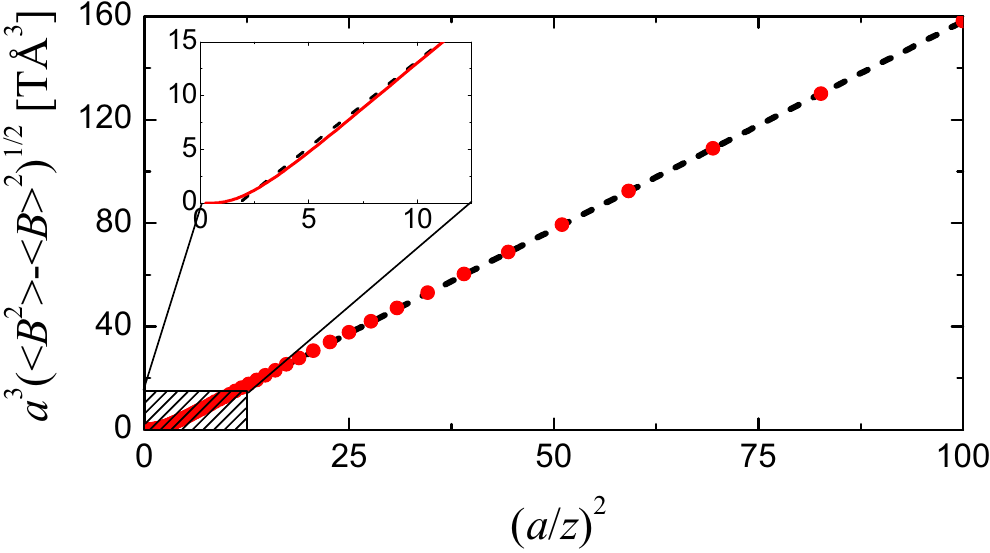}}
  \caption{The square root of the second moment of the field
    distribution as a function of distance from the monolayer (circles
    and solid line in inset). The inset is an expanded view of the
    shaded area. The dashed line is the fit described in the text.}
  \label{SecMom}
\end{minipage}
\end{figure}

In a LE-$\mu$SR or $\beta$-NMR measurements, the implanted probes stop
in the sample at a distribution of depths, $D(z)$, rather than a
unique single depth. This distribution can be simulated accurately
\cite{Morenzoni02NIMB,Keeler08PRB} using Trim.SP Monte-Carlo code
\cite{TRIM}. Therefore, the calculated $n(z,B)$ should be averaged
over $D(z)$. Moreover, since the probe stops in the substrate of the
monolayer, its intrinsic line shape (field distribution) has to be
convoluted with the depth averaged field distribution. Finally,
disorder in the monolayer, e.g.  deviations from a perfect triangular
lattice, can be taken into account by convoluting the depth averaged
distribution with a Gaussian of width that represents the degree of
disorder. In practice, one can simply convolute the calculated
distribution with a single Gaussian or Lorentzian that includes both
substrate and disorder effects. The resulting field distribution is
then
\begin{equation} 
  L(B)=I(B) * \int_{z_0}^{\infty} D(z) n(z,B) dz,
\end{equation}
where $I(B)$ is the broadening due to the substrate and disorder, and
$z_0$ is the distance between the magnetic cores in the monolayer and
the surface of the substrate. In the case of SMMs, $z_0$ is due to the
ligands used to graft the magnetic corse of a SMM to the substrate. In
Fig.~\ref{Measured}, we plot a typical field distribution due to a
monolayer of $1\mu_B$ moments, with $a \sim 5$ nm, $z_0 \sim 1$ nm and
a stopping profile as shown in the inset.
Both, the distribution with (solid line) and without (dashed line) the
broadening $I(B)$ are shown. Surprisingly, the asymmetry of the
distribution after averaging over depth is opposite to that of a
single depth, i.e. we find a low field instead of the high field tail
in Fig.~\ref{SumDip}. This is due to a larger near-zero field
contribution of probes implanted deeper into the substrate, far away
from the magnetic cores in the monolayer.  Note that unlike
Fig.~\ref{SumDip}, the most probable field in $L(B)$ is much closer to
$B=0$, and therefore in a real measurement we do not expect to see a
considerable shift in the precession frequency of the probe, instead
only some broadening can be measured \cite{Salman07NL,Salman09ML}.

Now we turn to the depth dependence of the field distribution. As we
mentioned earlier, we expect that for a depth $z \gg a$ the dipolar
field of the magnetic moment vanishes as in the case of a uniformly
magnetized infinite sheet. The depth dependence can be very useful in
measurements as a tool for identifying (I) the nature of the sensed
magnetic fields by the probe, dipolar or other, and (II) to estimate
the characteristic length scale, $a$, in the monolayer
\cite{Salman07NL}. In Fig.~\ref{SecMom} we plot the second moment of
the field distribution as a function of depth. The dashed line is a
linear fit of the second moment as a function of $(a/z)^2$ in the
range 50-100.  Clearly, the second moment is inversely proportional to
$z^2$. This is in contrast to the $1/z^3$ dependence of the dipolar
field from a single moment. A very small deviation from the $(a/z)^2$
behaviour can be seen for $z \geq a/3$ (see inset of
Fig.~\ref{SecMom}). Note also that the second moment vanishes at $z
\sim a$, as expected.

\section{Information Drawn from an Experiment}
Considering actual measurements, in \lemusr\ we measure the time
dependence of the polarization of the implanted muons in an applied
transverese field. This is proportional to the Fourier transform of
field distribution sensed by the muons, so that the average precession
frequency is propotional to the average field and damping rate is
proportional to the width of field distribution. In a \bnmr\
experiment we measure the \Li\ nuclear magnetic resonance line which
directly proportional to the field distribution. In both cases it is,
in principle, possible to fit the measured spectra. However, the
asymmetry of the field distribution is quite small, and therefore hard
to detect within the uncertainty of the measurement. Usually it is
sufficient to approximately fit the data assuming a simple Lorentzian
field distribution. As we mentioned above, there is no detectable
shift in the average field as a function of depth and temperature. In
contrast, the width of the distribution depends strongly on these
parameters. In fact, the width is proportional to $\langle \mu^2
\rangle$. Therefore, its temperature dependence can be obtained by
measuring the temperature dependence of the field distribution width
at a fixed depth (implantation energy). From this one should subtract
the additional broadening due to the substrate. This can be easily
evaluated by a measurement of the field distribution as a function of
temperature dependence at large depths (high implantation energy).
Finally, we point out that since we are dealing with local probe
measurements, the averaging in $\langle \mu^2 \rangle$, is on
individual moments during the lifetime of the probe. This is different
from the case of a XMCD measurement, where the averaging is done on
the absorption signal from all molecules.

\section{Conclusions}
We have presented a detailed calculation of the properties of the
magnetic field distribution, measured in the substrate, due to a
monolayer near its surface. This distribution and its depth dependence
can be used to infer the properties of the monolayer, such as the size
of magnetic moment and the average distance between neighbouring
moments. Although we restricted our calculations to oriented moments
on a triangular lattice, they can be easily generalized for different
systems.  Nevertheless, the depth dependence of the width of the
distribution (i.e. relaxation rate of the muon spin precession) and
its temperature and depth dependencies should not be affected by the
geometric details of the lattice.

\section*{References}
\bibliographystyle{elsarticle-num}
\newcommand{\noopsort}[1]{} \newcommand{\printfirst}[2]{#1}
  \newcommand{\singleletter}[1]{#1} \newcommand{\switchargs}[2]{#2#1}

\end{document}